\documentclass[12]{article}
\usepackage{fleqn}
\usepackage{graphicx}
\begin{document}
\Large
\begin{center}
{\bf Finite Projective Spaces, Geometric Spreads of Lines and Multi-Qubits}
\end{center}
\large
\vspace*{-.2cm}
\begin{center}
Metod Saniga
\end{center}
\vspace*{-.6cm}
\normalsize
\begin{center}
Astronomical Institute, Slovak Academy of Sciences\\
SK-05960 Tatransk\' a Lomnica, Slovak Republic

\vspace*{.2cm}

(6 July 2012)

\end{center}

\vspace*{-.3cm} \noindent \hrulefill

\vspace*{.0cm} \noindent {\bf Abstract}

\noindent
Given a $(2N - 1)$-dimensional projective space over GF(2), PG$(2N - 1,2)$, and its geometric spread of lines, there exists a remarkable mapping of this space onto PG$(N - 1,4)$ where the lines of the spread correspond to the points and subspaces spanned by pairs of lines to the lines of PG$(N - 1,4)$. Under such mapping, a non-degenerate quadric surface of the former space has for its image a non-singular Hermitian variety in the latter space, this quadric being {\it hyperbolic} or {\it elliptic} in dependence on $N$ being {\it even} or {\it odd}, respectively. We employ this property to show that generalized Pauli groups of $N$-qubits also form two distinct families according to the parity of $N$ and to put the role of symmetric Pauli operators into a new perspective.  The $N=4$ case is taken to illustrate the issue, due to its link with the so-called black-hole/qubit correspondence.

\vspace*{-.2cm} \noindent \hrulefill

\vspace*{.4cm}

Multiple qubit states play a key role in various fields of quantum information theory like quantum computing, coding and quantum error-correction (see, e.\,g., \cite{n-c}). Recently, and rather surprisingly, they have also been recognized to be of great relevance for getting insights into the nature of
entropy formulas of a certain class of stringy black hole solutions (see, e.\,g., \cite{duff}). It is, therefore, important to deepen our understanding of these fundamental buildings blocks of quantum world.  In the present note we do so through the geometry of their associated generalized Pauli groups.

Let PG$(d,q)$ be a $d$-dimensional projective space over GF$(q)$, $q$ being a power of a prime.\footnote{For the standard mathematical nomenclature and notation employed in what follows, see, e.\,g., \cite{h-t}.} A $t$-spread $\mathcal{S}$ of  PG$(d,q)$ is a set of $t$-dimensional subspaces of PG$(d,q)$ which partitions its point-set \cite{ei-st}. If the elements of $\mathcal{S}$ in a subspace $V$ form a $t$-spread on $V$, one says that $\mathcal{S}$ induces a $t$-spread on $V$. A $t$-spread $\mathcal{S}$ is called {\it geometric} (or normal) if it induces a $t$-spread on each $(2t+1)$-dimensional subspaces of PG$(d,q)$ spanned by a pair of its elements. It is a well-known fact that PG$(d,q)$ possesses a $t$-spread iff $(t+1)|(d+1)$; moreover, this condition is also sufficient for PG$(d,q)$ to have a geometric $t$-spread. B. Segre showed \cite{segre} that a geometric $t$-spread  of PG$(N(t+1) - 1,q)$, $N \geq 2$, gives rise to a projective space
PG$(N - 1,q^{t+1})$ as follows: the points of this space are the elements of $\mathcal{S}$ and its lines are the $(2t+1)$-dimensional subspaces spanned by any two distinct elements of $\mathcal{S}$, with incidence inherited from PG$(N(t+1) - 1,q)$. For a particular case of $t=1$ (i.\,e., a spread of lines), Dye \cite{dye} demonstrated that a {\it hyperbolic} or an {\it elliptic} quadric of PG$(2N -1,q)$ has an induced (geometric) spread of lines if and only if {\it N} is, respectively, {\it even} or {\it odd}, in which case it is mapped onto a non-singular Hermitian variety H$(N-1,q^{2})$ of PG$(N-1,q^{2})$. We shall now show that this  property has for $q=2$ a very interesting physical implication.

It is already a firmly established fact \cite{sp,hos,thas,v-l} that the commutation relations between the elements of the generalized Pauli group of $N$-qubits,  $N \geq 2$, can be completely reformulated in the geometrical language of symplectic polar space of rank $N$ and order two, W$(2N-1,2)$; the generalized Pauli operators (discarding the identity)
answer to the points of W$(2N-1,2)$, a maximally commuting subset has its representative in a maximal totally isotropic subspace of
W$(2N-1,2)$ and commuting translates into collinear. One of the most natural representations of W$(2N-1,2)$ is that in terms of the points and the set of totally isotropic subspaces of PG$(2N-1,2)$ endowed with a symplectic polarity. Employing this representation, it has been found in \cite{hos} that in the real case the {\it symmetric} elements/operators of the $N$-qubit Pauli group {\it always} lie on a {\it hyperbolic} quadric in the ambient space PG$(2N-1,2)$. Combining this fact with Dye's result, we arrive at our main observation: {\it it is only for $N$ \underline{even} when \underline{all} symmetric generalized Pauli operators of W$(2N-1,2)$  can be mapped to the points of an Hermitian variety of the space PG$(N - 1,4)$ associated through a geometric spread of lines with the ambient space PG$(2N-1,2)$.} Hence, in this regard, when it comes to generalized Pauli groups `even-numbered' multi-qubits are found to stand on a slightly different footing than `odd-numbered' ones.

We shall finish this communication by briefly mentioning an especially  interesting even case, $N=4$. Here, a hyperbolic quadric Q$^{+}(7,2)$ of PG$(7,2)$ formed by the symmetric operators is well known for its puzzling triality swapping points and two systems of generators  and has for its spread-induced image  an Hermitian surface H$(3,4)$ of PG$(3,4)$ (see, e.\,g., \cite{hos2}). This Hermitian surface is, in turn, nothing but the generalized quadrangle GQ$(4,2)$ in disguise (see, e.\,g., \cite{pay-thas}), the dual of which --- GQ$(2,4)$ --- was found to play a prominent role in the so-called black-hole-qubit correspondence, by fully encoding the $E_{6(6)}$ symmetric entropy formula describing black holes and black strings in $D=5$ \cite{gq24}. Our finding thus, {\it inter alia}, not only opens up an unexpected window through which also four-qubit Pauli group, like its lower rank cousins, could find its way into some black hole entropy formula(s), but also puts the role of symmetric operators into a new perspective. It is also important to keep in mind this remarkable {\it three-to-one} correspondence, i.\,e., that it is always a triple of (collinear) operators of the ambient space PG$(7,2)$ which comprises a single point of PG$(3,4)$.

\normalsize
\section*{Acknowledgements}
This work was partially supported by the VEGA grant agency, projects Nos. 2/0092/09 and 2/0098/10. The idea exposed in this paper originated from discussions with Prof. Hans Havlicek, Dr. Boris Odehnal (Vienna University of Technology) and Dr. Petr Pracna (J. Heyrovsk\' y Institute of Physical Chemistry, Prague).

\end{document}